\documentclass[aps,twocolumn,groupedaddress,showpacs]{revtex4-1}
\usepackage{epstopdf}
\usepackage{physics}
\usepackage{tikz,lipsum,lmodern}
\usepackage[export]{adjustbox}
\usepackage{subcaption}
\usepackage{amsmath}
\usepackage{amssymb}
\usepackage{svg}

\usepackage[unicode=true,pdfusetitle,
 bookmarks=true,bookmarksnumbered=false,bookmarksopen=false,
 breaklinks=false,pdfborder={0 0 0},backref=false,colorlinks=true,citecolor=blue,urlcolor=violet 
]{hyperref}
\usepackage{xcolor}
\usepackage{graphicx}
\newcommand{\orcid}[1]{\href{https://orcid.org/#1}{\includegraphics[width=10pt]{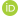}}}
\usepackage{setspace}
\begin{document}
\title{Coherence, Path-Predictability and I-Concurrence: A Triality}

\author{Abhinash Kumar Roy$^1$\orcid{0000-0001-7156-1989}}
\email{akr16ms137@iiserkol.ac.in}
\author{Neha Pathania$^2$\orcid{0000-0002-3385-4761}}
\email{neha@ctp-jamia.res.in}
\author{Nitish Kumar Chandra$^1$\orcid{0000-0002-6572-1322}}
\email{nitishkrchandra@gmail.com}
\author{Prasanta K. Panigrahi$^1$\orcid{0000-0001-5812-0353}}
\email{pprasanta@iiserkol.ac.in}
\author{Tabish Qureshi$^2$\orcid{0000-0002-8452-1078}}
\email{tabish@ctp-jamia.res.in}

\affiliation{$^1$Department of Physical Sciences, Indian Institute of Science Education and Research Kolkata, Mohanpur, West Bengal-741246, India}
\affiliation{$^2$Centre for Theoretical Physics, Jamia Millia Islamia, New Delhi-110025, India}
\begin{abstract}
It is well known that fringe contrast is not a good quantifier of the wave
nature of a quanton in multipath interference. A new interference visibility,
based on the Hilbert-Schmidt coherence is introduced. It is demonstrated that
this visibility is a good quantifier of wave nature, and can be experimentally
measured. A generalized path predictability is introduced, which reduces to
the predictability of Greenberger and Yasin, for the case of two paths.
In a multipath, which-way interference experiment, the new visibility,
the predictability and the I-concurrence (quantifying the entanglement
between the quanton and the path-detector), are shown to follow a
tight \emph{triality} relation. It \emph{quantifies} the essential role that
entanglement plays in multipath quantum complementarity, for the first time.
\end{abstract}

\maketitle

\section{Introduction}

We have come a long way since Neils Bohr proposed his principle of 
complementarity in 1928 \cite{bohr}, but instead of fading away, the topic
has increasingly become more interesting and involved. Wootters and Zurek
were the first ones to quantitatively analyze the phenomenon of complementarity 
or wave-particle duality, as it is more commonly called \cite{wootters}.
Greenberger and Yasin analyzed the problem with the assumption that unequal
beams in a two-path interferometer allows for predicting, to a certain
degree, which of the two paths the quanton followed. They established
a duality relation \cite{GY}
\begin{equation}\label{dualitygy}
   {P}^{2} + {V}^{2} \le  1,
\end{equation}
which involved a \emph{path predictability} ${P}$ and an interference
visibility, which is just the fringe contrast
$V=(I_{max}-I_{min})/(I_{max}+I_{min})$, $I_{max},I_{min}$
being the maximum and minimum intensities in a region on the screen. Here, equal beams would imply no path information.
It was further refined by Jaeger, Shimony and Vaidman \cite{jaeger}.
This approach, however, does not take into account the fact that even when the
two beams are equal, one may introduce a probe to find out which of
the two paths the quanton followed. Such a scenario was considered by
Englert \cite{englert}, where he studied a two-path interferometer in the presence of
a path-detector. He arrived at a duality relation 
\begin{equation}\label{dualitye}
   {D}^{2} + {V}^{2} \le  1,
\end{equation}
involving a \emph{path distinguishability} ${D}$ and the fringe 
contrast ${V}$. After a lull of nearly two decades, the issue
of wave-particle duality saw a resurgence when the concept of
complementarity was quantitatively extended to mulitpath interference
\cite{cd15,nduality,bagan}. The multipath complementarity relations used
a \emph{coherence} introduced in the context of quantum information
theory \cite{plenio}, instead of conventional interference visibility.
This coherence measure has been shown to be a good quantifier of the wave
nature of a quanton \cite{coherence}.
The duality relation, based on path-predictability, was also generalized
to the multipath case using coherence \cite{roy}. However, the two kinds of duality
relations, one based on predictability, and the other on distinguishability,
remained separated.

\begin{figure}
\centerline{\resizebox{8.0cm}{!}{\includegraphics{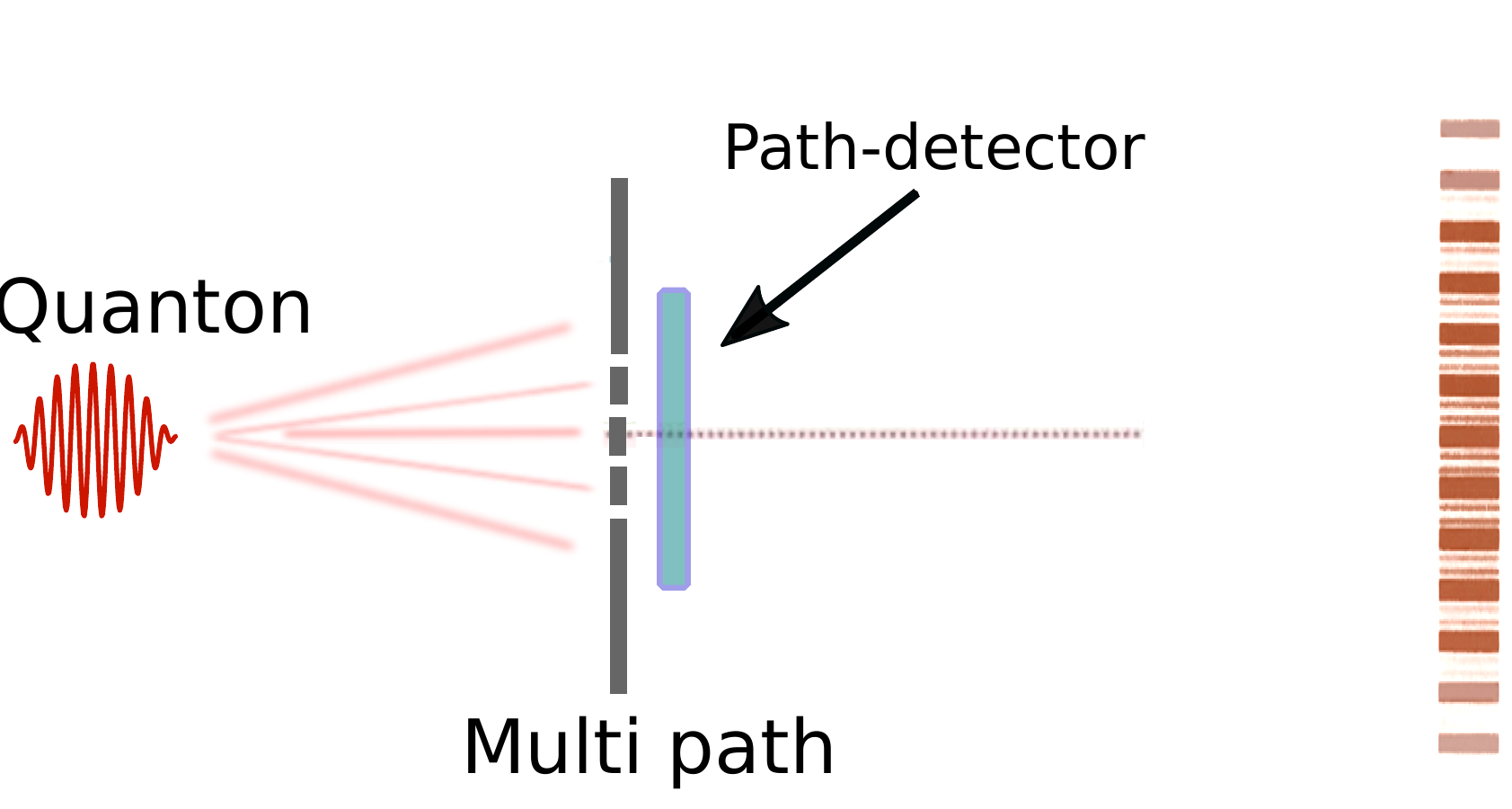}}}
\caption{A schematic representation of a multipath interference experiment, with which-way detection. 
}
\label{npath}
\end{figure}

Jakob and Bergou \cite{triality1} were the first to point out that, in
two-path interference,
the predictability and distinguishability were connected through
entanglement, and the duality relation (\ref{dualitye}) can be written as
a triality relation between predictability, visibility and concurrence
\cite{concurrence}, which is a measure of entanglement. This was followed
by a similar realization in classical optics \cite{triality2}.
This idea was later investigated in a multipath quantum setting,
and a multipath predictability, coherence \cite{plenio} and entanglement
were shown to follow a triality relation \cite{triality3}. While that
analysis used a robust measure of coherence, the quantifier of
entanglement could not be shown to be a proper entanglement measure.
Although the analysis brought out the role of entanglement in multipath
complementarity, it could not be rigorously quantified. There have been other interesting works trying to find an operational entanglement measure from complementarity in the general case of multipartite and multi-dimensional systems \cite{huber}.

In the following, we investigate the issue of complementarity in multipath
interference and show that a robust entanglement measure, namely
I-concurrence \cite{rungta}, plays an essential role in it.

\section{Preliminaries}

\subsection{Visibility measure}

Two decades back it was experimentally demonstrated that in multipath
interference, conventional visibility, defined as fringe contrast, can
\emph{increase} under path-selective decoherence or selective path
information \cite{mei}. Subsequent debate concluded that conventional
visibility does not quantify the wave nature of a quanton in multipath
interference, and cannot be used to analyze the notion of Bohr's complementarity
\cite{luis,bimonte1,prillwitz}. In the experiment of Mei and Weitz \cite{mei}, 
in a multipath interference, one particular path was given an additional
180$^\circ$ phase flip. A selective decoherence was applied on the particular
path, by scattering photons off the quanton. Counter-intutively, this resulted in an increase in the fringe contrast of the interference pattern.

We use the normalized Hilbert-Schmidt coherence to define a generalized visibility for a quanton passing through a multipath interferometer (see FIG. \ref{npath})
\begin{equation}
\mathcal{V}^{2} \equiv \frac{n}{n-1}\sum_{i\neq j} |\rho_{ij}|^{2} = \mathcal{C}_{HS} ,
\label{VHS}
\end{equation}
where $\mathcal{C}_{HS}$ is the normalised Hilbert-Schmidt coherence. The system is assumed to have a $n$-dimensional Hilbert space, which arises out of the quanton passing through $n$ paths. The states corresponding to the quanton following each path are mutually orthogonal, and can be used to construct a Hilbert space basis. The density matrix element $\rho_{ij}$ is taken between two such states. This visibility 
satisfies all the criteria proposed by D{\"u}rr for a valid measure of
interference and wave properties \cite{durr}. 
It is worth noting that, for the two slit interference case, the
generalized visibility reduces to $\mathcal{V} = 2|\rho_{12}|$, which can be easily
shown to be equal to the usual visibility measure ${V} = (I_{max}
- I_{min})/(I_{max} + I_{min})$.

It might be useful to mention at this point that in classical optics, the interference visibility in a two-slit interference is equal to the degree of mutual coherence between the two slits, for equal intensity in the two beams. Thus for a two-slit experiment, $2\rho_{12}$ can be identified with mutual coherence function $\gamma_{12}$, at least for equal intensity in the two beams. The classical optical equivalent of the Hilbert-Schmidt coherence then would be proportional to the square of mutual coherence function between two slits, summed over all slit pairs, $\sum_{i\neq j} |\gamma_{ij}|^2$.

Consider a quanton passing through $n$-path interferometer, which could
have a multi-slit or some kind of beam splitter, as the first element.
The state corresponding to the $i^{th}$ path can be represented by the pure
state $|\psi_i\rangle$. In general the density operator of the quanton
may be mixed, and will have the general form
\begin{equation}
  \rho = \sum_{j=1}^n \sum_{k=1}^n \rho_{jk} |\psi_j\rangle\langle\psi_k|.
\end{equation}
The diagonal elements represent the fractional populations of the various beams.
We allow the possibility that the phase in $i^{th}$ beam may be shifted by
$\theta_i$, in order to analyze the experimental scenario of Mei and Weitz
\cite{mei}.
On coming out, the state of the quanton is
\begin{equation}
  \rho = \sum_{j=1}^n \sum_{k=1}^n \rho_{jk} |\psi_j\rangle\langle\psi_k|
  e^{i(\theta_j-\theta_k)},
\end{equation}
where $\rho_{jk}$ are assumed to be real without loss of generality, as any phases can be absorbed in the exponential factor.
In the multi-slit interference experiment, after coming out, the beams spread
and overlap and are registered at different positions on the screen. In another
kind of multipath interference, the beams may be split and recombined,
as happens in a Mach-Zehnder interferometer. In general, we may assume that
each beam is split into new channels, whose states may be represented
by $|\xi_i\rangle$. The simplest case would be where
all the original beams have equal overlap with a
particular output channel, say $|\xi_i\rangle$. This translates to an
assumption that $\langle\xi_i|\psi_k\rangle = A_i$ is same for all $k$.
The probability $I$ of finding the quanton in the $i^{th}$ output channel
(not the $i^{th}$ path) is then
given by $I = \langle\xi_i|\rho|\xi_i\rangle$. For the present case,
this probability is given by
\begin{eqnarray}
I &=& |A_i|^2 \Big[ \sum_{j=1}^n \rho_{jj} +
\sum_{j \neq k}\rho_{jk} e^{i(\theta_j-\theta_k)} \Big] \nonumber\\
  &=& |A_i|^2 \Big[1 + 
\sum_{j \neq k}\rho_{jk} e^{i(\theta_j-\theta_k)} \Big] \nonumber\\
  &=& |A_i|^2 \Big[1 + 
\sum_{j \neq k}|\rho_{jk}| \cos(\theta_j-\theta_k) \Big] .
\label{ninterf}
\end{eqnarray}
If one were to simulate Mei and Weitz experiment in this situation,
one would first change the phase of the $\ell$'th path as
$\theta_{\ell} \rightarrow \theta_{\ell}+\pi$. It is to be noted that our visibility $\mathcal{V}$, given by (\ref{VHS}), is unaffected
by this phase change, whereas the fringe contrast does get affected.

If one wants to know which path the quanton followed, one needs to have
the quanton paths entangled with a path-detector.
The combined state of the quanton and the path-detector, after the two
have interacted, will be of the form
\begin{equation}
  \rho' = \sum_{j=1}^n \sum_{k=1}^n \rho_{jk} |\psi_j\rangle\langle\psi_k|
  e^{i(\theta_j-\theta_k)}\otimes |d_j\rangle\langle d_k|,
\end{equation}
where $\{|d_i\rangle\} $ are the states of the path detector, assumed to be
normalized, but not necessarily mutually orthogonal.
The reduced density operator of the quanton is obtained by tracing over the
path-detector states:
\begin{equation}
  \rho_r' = \sum_{j=1}^n \sum_{k=1}^n \rho_{jk} |\psi_j\rangle\langle\psi_k|
  e^{i(\theta_j-\theta_k)} \langle d_k|d_j\rangle .
\end{equation}
In order to reproduce the Mei and Weitz experimental situation, we assume
that the path-detector can only tell if the quanton passed through path
$\ell$ or not, it cannot discriminate between any other paths. This is 
achieved by having all the path detector states being identical,
(say) equal to $|d_1\rangle$, and only $|d_{\ell}\rangle$ being different
from $|d_1\rangle$. Now if one were to calculate the visibility, as
defined by (\ref{VHS}), for this state of the quanton, it will turn out
to be
\begin{equation}
\mathcal{V}'^{2} = \frac{n}{n-1}\sum_{i\neq j} |\rho_{ij}|^{2}
|\langle d_k|d_j\rangle|^2 .
\label{VHS'}
\end{equation}
Since $|\langle d_i|d_j\rangle|=1$ for all $i,j$ except those involving $\ell$,
one can easily see that
\begin{equation}
\mathcal{V}'^{2} \neq \frac{n}{n-1}\sum_{i\leq j} |\rho_{ij}|^{2}
\end{equation}
or
\begin{equation}
\mathcal{V}'^{2} \leq \mathcal{V}^{2} ,
\end{equation}
which means that the visibility will never increase (equality is achieved for the trivial case when visibility is zero i.e., for diagonal quanton states) if one tries to find out
if the quanton passed through path $\ell$ or not . Thus the new visibility 
works satisfactorily even for Mei and Weitz experiment, where conventional
visibility fails.

The next question we address is, how to measure this visibility in an
interference experiment. The probability given by (\ref{ninterf}), can
also be interpreted as the intensity at a position on the screen, or in
an output channel. In general it depends on the phases of the paths.
D\"urr has shown that the second moment of this interference pattern \cite{durr},
\begin{equation}
\langle (\Delta I)^2\rangle_{\theta} = \tfrac{1}{n^2}
\sum_{i\neq j} |\rho_{ij}|^{2} ,
\end{equation}
where $\Delta I = I - \langle I\rangle_{\theta}$, and
$\langle\dots \rangle_{\theta}$ denotes the average over all phases.
Thus the new visibility can be written as
\begin{equation}
\mathcal{V}^{2} = \tfrac{n^3}{n-1} \langle (\Delta I)^2\rangle_{\theta},
\end{equation}
which means, it can be measured if one can design an experimental
arrangement to vary the phases of all paths independently, and do an averaging
over them. This is one method of experimentally obtaining the new visibility.

Notice that (\ref{VHS}) represents a sum over pairs of paths. If one were
to modify the $n$-path interference experiment by blocking all paths except
one pair $i,j$, the \emph{conventional visibility} of the interference of
that pairs of paths will be given by
\begin{equation}
{V}_{ij} = \frac{2|\rho_{ij}|}{\rho_{ii}+\rho_{jj}}.
\end{equation}
Now one can repeat the experiment with another pair of paths, and so on, until
one has gone over all path pairs. If all paths are equally probable, 
$\rho_{ii}=\rho_{jj}=\frac{1}{n}$, ${V}_{ij} = n|\rho_{ij}|$.
The \emph{mean square} of the visibilities, of all path pairs, is given by
\begin{equation}
\tfrac{2}{n(n-1)} \sum_{\text{pairs}}{V}_{ij}^2 = 
\tfrac{n}{n-1}\sum_{i\neq j}|\rho_{ij}|^2 = \mathcal{V}^2 .
\end{equation}
This is a very interesting result, which shows that the multipath visibility
$\mathcal{V}$ is the \emph{root mean square} of the fringe contrasts of 
interference from all path pairs. So, another way $\mathcal{V}$ can
be measured is by opening only a pair of paths at a time, and measuring
conventional visibility, and then taking root mean square over all path
pairs. Such a procedure is very much feasible, and selectively opening
path pairs has already been experimentally demonstrated \cite{padua}.
If the paths are not equally probable, each two-path visibility has to be multiplied with $\frac{n}{2}(\rho_{ii}+\rho_{jj})$ to normalize the weights before taking the average. It can be shown that 
$\tfrac{n}{2(n-1)} \sum_{\text{pairs}}(\rho_{ii}+\rho_{jj})^2{V}_{ij}^2 = 
\ \mathcal{V}^2$.

\subsection{Predictability measure}
A generalized predictability can be defined as,
\begin{equation}\label{pred1}
  \mathcal{P}^{2}   = \sum_{i =1}^{n}\rho_{ii}^{2} - \frac{1}{n-1}\sum_{i\neq j}\rho_{ii}\rho_{jj} = 1 - \frac{n}{n-1}\sum_{i\neq j}\rho_{ii}\rho_{jj}
\end{equation}
where $\rho$ is the density matrix of the quanton and $n$ is the
dimension of the density matrix (number of paths). As is evident from
the above expression $\mathcal{P}$ varies from zero to one, therefore
normalized. Although defined in a very different manner, it turns out that
(\ref{pred1}) is identical to the predictability defined by D\"urr \cite{durr}.
To show that (\ref{pred1}) is a good measure of generalized
predictability, we show that it follows all of D{\"u}rr's criteria.\\
(i) $\mathcal{P}$ is a polynomial function of $\rho_{ii}$, therefore
a continuous function involving only the diagonal elements.\\ (ii)
When the quanton's path is known for sure, i.e., $\rho_{ii} = 1$ and
$\rho_{jj} = 0~~ \forall ~~j \neq i$ (only $i^{th}$ slit is open),
predictability  reaches its global maximum i.e., $\mathcal{P} =
1$.\\ (iii) When quanton's probability of going through all paths
are equally likely, i.e., $\rho_{ii} = \frac{1}{n}~~\forall~~i$,
generalized predictability is zero, since $\mathcal{P}^{2} = 1 -
\frac{n}{n-1}\sum_{i\neq j}\frac{1}{n^{2}} = 0$.\\ (iv) A change of
probabilities (diagonal elements $\rho_{ii}$) towards equalization leads
to decrease in the generalized predictability. To show that, consider
$\rho_{11} < \rho_{22}$ and a small parameter $0<\varepsilon < \rho_{22}
- \rho_{11}$. A shift in probabilities, $\Tilde{\rho}_{11} = \rho_{11}
+ \varepsilon$ and $\Tilde{\rho}_{22} = \rho_{22} - \varepsilon$, with
other elements unchanged yields the following,
\begin{equation}
    \mathcal{P}^{2} - \Tilde{\mathcal{P}}^{2} = \frac{2n\varepsilon}{n-1}\left(\rho_{22} - \rho_{11} -\varepsilon\right).
\end{equation}
Since, $\varepsilon>0$ and $\varepsilon < \rho_{22} - \rho_{11}$, one obtains $\mathcal{P} > \Tilde{\mathcal{P}}$, therefore, a decrease in predictability.

Therefore, generalized predictability as defined in (\ref{pred1}) satisfies D{\"u}rr's criteria. Furthermore, for two beam interferometer case ($n=2$), it reduces to,
\begin{equation}
    \mathcal{P} = |\rho_{11} - \rho_{22}|
\end{equation}
which matches with the predictability proposed by Greenberger and Yasin
\cite{GY}.

Interestingly, this multipath predictability can also be written as root-mean-square of the two-path predictabilities from all path pairs:
\begin{equation}
    \mathcal{P}^2 = \tfrac{1}{n-1} \sum_{\text{pairs}}(\rho_{ii}+\rho_{jj})^2{P}_{ij}^2,
\end{equation}
where ${P}_{ij}$ is the path predictability for the i'th and j'th path, \emph{a la} Greenberger and Yasin.
The fact that our $n$-path predictability can also be expressed in terms of the conventional two-slit predictability for path pairs, has not been recognized earlier, although the predictabilities proposed by Greenberger
and Yasin \cite{GY} and D\"urr \cite{durr} have been around for a long time. 

\subsection{Path distinguishability}
A multipath distinguishability has been proposed earlier as \cite{nduality}  
\begin{equation}\label{dis1}
      \mathcal{D}^{2} = 1 - \left(\frac{1}{n-1} \sum_{i\neq j}\sqrt{\rho_{ii}\rho_{jj}}|\langle d_{i}|d_{j}\rangle|\right)^{2},
\end{equation}
which can be related to unambiguous quantum state discrimination.
We define the path-distinguishability $\mathcal{D}$ in a different way: 
\begin{equation}\label{dis2}
    \mathcal{D}^{2} = 1 - \frac{n}{n-1} \sum_{i\neq j}\rho_{ii}\rho_{jj}|\langle d_{i}|d_{j}\rangle|^{2}.
\end{equation}
In the spirit of complementarity, our proposed distinguishability
measure satisfies the following basic criteria for being a reliable path
quantifier \cite{durr}:  \\(i) When the detector states $\{{|d_{i}\rangle}\}$ form
an orthonormal basis, i.e., $\langle d_{i}|d_{j}\rangle = \delta_{ij}$,
the path distinguishability $\mathcal{D}$ is maximum ($\mathcal{D} = 1$),
since in this case all the paths can be unambiguously discriminated.
\\  (ii) When all the detector states are parallel, i.e., state of
quanton and the detector is separable, and probability through all
the slits are equal, i.e., $\rho_{ii} = 1/n~\forall i$, then the path
distinguishability $\mathcal{D}$ is minimum ($\mathcal{D} = 0$). \\(iii)
Whenever the detector states $|d_{i}\rangle$ and $|d_{j}\rangle$ are made
more orthogonal i.e., with an increment in $|\langle d_{i}|d_{j}\rangle|$,
the distinguishability increases. \\(iv) Distinguishability and visibility
both can not increase simultaneously under any operation, which is evident
through the fact $\mathcal{D}^{2} + \mathcal{V}^{2} = 1$.

For the case of two paths, (\ref{dis2}) reduces to the distinguishability
defined by Englert \cite{englert}. Next we address the issue of measuring
this distinguishability. Taking cue from the visibility, we note that if 
one were to close all paths, except paths $i,j$, the path distinguishability,
based on minimum error state discrimination, is given by
\begin{equation}
{D}_{ij} = \sqrt{1 - \frac{4\rho_{ii}\rho_{jj}}{(\rho_{ii}+\rho_{jj})^2}
|\langle d_{i}|d_{j}\rangle|^{2}},
\end{equation}
which is essentially Englert's distinguishability, generalized for unequal path probabilities by using the Helstrom bound \cite{helstrom}.
As done in the case of visibility, one can repeat the interference experiment by selectively
opening different pairs of slits, while measuring the distinguishability. When all the path are equally probable, the
\emph{mean square} of the distinguishabilities, from all pairs of paths,
is given by
\begin{equation}\label{diseq}
\tfrac{2}{n(n-1)} \sum_{\text{pairs}}{D}_{ij}^2 = 
1 - \frac{n}{n-1} \sum_{i\neq j}\rho_{ii}\rho_{jj}|\langle d_{i}|d_{j}\rangle|^{2}
 = \mathcal{D}^2,
\end{equation}
which is the same as (\ref{dis2}). Thus the distinguishability too can be
measured experimentally, as the \emph{root mean square} of 
conventional distinguishabilities, from all path pairs. When the paths are not equally probable, the distinguishability can be measured through the conventional two-slit distinguishability from path pairs $D_{ij}$ and the path probabilities $\rho_{ii}$ as follows
\begin{eqnarray}\label{disuneq}
    \mathcal{D}^{2} &=& \frac{n}{2(n-1)}\sum_{pairs}(\rho_{ii} + \rho_{jj})^{2} D_{ij}^{2} \nonumber\\
    && + 1 - \frac{n}{2(n-1)}\sum_{pairs}(\rho_{ii} + \rho_{jj})^{2}.
\end{eqnarray}
The last two terms in the above relation may be interpreted as a measure of the inequality in the probability of various path pairs, by rewriting them as $1 - \frac{2}{n(n-1)}\sum_{pairs}\frac{n^2}{4}(\rho_{ii} + \rho_{jj})^{2}$. For equally probable paths $\rho_{ii}+\rho_{jj}=2/n$ for all $i,j$.
One can check that for equally probable paths, only the first term survives and reduces to (\ref{diseq}). For $n=2$, there is only one path pair, and there is no question of inequality between different pairs. In that case, the last two terms will be zero even if the two paths are unequally probable.

\section{Complementarity}
\subsection{Role of entanglement}
Consider a quanton passing through a $n$-path interferometer, such that its quantum state is given by the following pure state,
\begin{equation}\label{pure1}
    |\psi\rangle = \sum_{i=1}^{n}c_{i}|\psi_{i}\rangle,
\end{equation}
where, $c_{i}\in \mathbf{C}$. The density operator for the quanton can be written as $\rho = \sum_{i,j}c_{i}c_{j}^{*}|\psi_{i}\rangle\langle\psi_{j}|$.
Using (\ref{VHS}) and (\ref{pred1}),
one obtains a $n$-path duality relation as follows. The sum of the squares of the predictability and visibility is given by
\begin{eqnarray}\label{duality1}
    \mathcal{P}^{2} + \mathcal{V}^{2} &=& 1 - \frac{n}{n-1}\sum_{i\neq j}\left(\rho_{ii}\rho_{jj}-|\rho_{ij}|^{2}  \right) .
\end{eqnarray}
Since for the pure state given by (\ref{pure1}) $\rho_{ij}
= c_{i}c_{j}^{*}$ resulting in $\rho_{ii}\rho_{jj} =
|\rho_{ij}|^{2}$, the duality relation saturates
for the pure state case. When a quanton is given by an ensemble
$\{(p^{\alpha},~ |\psi^{\alpha}\rangle)\}$, where $p^{\alpha}
\geq 0$ with $\sum p^{\alpha} = 1$, and $|\psi^{\alpha}\rangle =
\sum_{i}c^{\alpha}_{i}|\psi_{i}\rangle$. The resulting density
matrix is mixed with the elements given by $\rho_{ij} =
\sum_{\alpha}p^{\alpha}c_{i}^{\alpha}c_{j}^{\alpha *}$ :
\begin{equation}
    \rho_{ii}\rho_{jj}-|\rho_{ij}|^{2} = \sum_{\alpha}\left\lvert c_{i}^{\alpha}\right\rvert^{2} \sum_{\beta}\lvert c_{j}^{\beta}\rvert^{2} - \abs{\sum_{\alpha}c_{i}^{\alpha}c_{j}^{\beta *}}^{2} \geq 0.
\end{equation}
Here, the last inequality follows from the Cauchy-Schwartz inequality. Therefore, in general we have the following $n$-path duality relation,
\begin{equation}
    \mathcal{P}^{2} + \mathcal{V}^{2} \leq 1,
\end{equation}
which saturates for pure $\rho$. This result was already derived by 
D\"urr \cite{durr}, but now sets the stage for investigating the case where
there is a path-detector in place, which can perform a which-path or
which-way detection.

Consider an interferometer with an added path-detecting device, such that the
quanton going through the $n$ paths is described by the state
\begin{equation}
    |\Psi\rangle = \sum_{i=1}^{n}c_{i}|\psi_{i}\rangle|d_{i}\rangle,
\end{equation}
where, $|\psi_{i}\rangle$ is the state of the quanton corresponding to its passing through the $i^{th}$ path, and $|d_{i}\rangle$ is the state of the path-detector, corresponding to that possibility. The states $\{|d_{i}\rangle\}$ are assumed to be normalized, but not necessarily orthogonal.

The reduced density operator of the quanton, after tracing over the path-detector states is, 
\begin{equation}\label{reduceddensity1}
    \rho_{r} = \sum_{i,j}c_{i}c_{j}^{*}\langle d_{j}|d_{i}\rangle~|\psi_{i}\rangle\langle \psi_{j}|.
\end{equation}
In terms of the density operator for the quanton in the absence of the path-detector, one can write  (\ref{reduceddensity1}) as,
\begin{equation}
    \rho_{r} = \sum_{i,j}\rho_{ij}\langle d_{j}|d_{i}\rangle~|\psi_{i}\rangle\langle \psi_{j}|
\end{equation}
 For this reduced density operator, one can write (\ref{duality1}) as,
 \begin{equation}\label{duality2}
 \begin{aligned}
     \mathcal{P}^{2} + \mathcal{V}^{2} =& 1 - \frac{n}{n-1}\sum_{i\neq j}\rho_{ii}\rho_{jj}+\frac{n}{n-1}\sum_{i\neq j}|\rho_{ij}|^{2}|\langle d_{j}|d_{i}\rangle|^{2}\\ =& 1 - \frac{n}{n-1}\sum_{i\neq j}\rho_{r_{ii}}\rho_{r_{jj}}+\frac{n}{n-1}\sum_{i\neq j}|\rho_{r_{ij}}|^{2} 
     \end{aligned}
 \end{equation}
One can define a normalized entanglement measure $\mathcal{E}$ such that,
\begin{equation}
    \mathcal{E}^{2} = \frac{n}{n-1}\sum_{i\neq j}\left(\rho_{r_{ii}}\rho_{r_{jj}} - |\rho_{r_{ij}}|^{2} \right) = \frac{n}{2(n-1)}E^{2}
\end{equation}
where $E$ is the generalized concurrence \cite{gconc}, and coincides
with I-concurrence \cite{rungta}, when the bipartite state under consideration is a pure state. Eqn. (\ref{duality2}) can then be written as,
\begin{equation}
    \mathcal{P}^{2} + \mathcal{V}^{2} = 1 - \mathcal{E}^{2},
\end{equation}
or as a triality relation,
\begin{equation}\label{triality}
     \mathcal{P}^{2} + \mathcal{V}^{2} + \mathcal{E}^{2}= 1 .
\end{equation}

Therefore, the generalized predictability, Hilbert-Schmidt coherence,
and I-concurrence obey a tight triality relation. It is interesting
to note that the path-distinguishability measure, as defined in the
previous section, is related to the generalized predictability through
I-concurrence as follows,
\begin{equation}\label{DPE1}
   \mathcal{D}^{2} =  \mathcal{P}^{2} + \mathcal{E}^{2} .
\end{equation}
This above relation shows that the two different kinds of path knowledge,
predictability and distinguishability, are quantitatively connected through 
a measure of entanglement, the I-concurrence. It very naturally generalizes 
Englert's duality relation (\ref{dualitye}) to multipath interference:
\begin{equation}\label{nduality}
   \mathcal{D}^{2} + \mathcal{V}^{2} =  1.
\end{equation}
Needless to say, Englert's duality relation is recovered for $n=2$.
Since the above relation is an equality, it is obvious that this
multipath distinguishability does not suffer from the shortcoming of
D\"urr's multipath distinguishability, pointed out by Bimonte and
Musto \cite{bimonte2}.

We have shown that a proper 
entanglement measure, the I-concurrence, is an integral part of complementarity
in multipath interference. For $n=2$, (\ref{triality}) reduces to
\begin{equation}\label{triality2}
     {P}^{2} + {V}^{2} + \mathcal{E}_2^{2}= 1 ,
\end{equation}
where is ${P}$ is the predictability of Greenberger and Yasin \cite{GY}, ${V}$ is
the fringe contrast, and $\mathcal{E}_2$ is the concurrence \cite{concurrence}.
This two-path result is the same as that derived earlier by Jakob and
Bergou \cite{triality1}, and it is satisfying to see that our general
multipath relation reduces to it in the appropriate limit. In addition,
for $n=2$ the relation (\ref{DPE1}) reduces to 
\begin{equation}\label{DPE}
   {D}^{2} =  {P}^{2} + \mathcal{E}_2^{2} ,
\end{equation}
where ${D}^{2}$ is Englert's distinguishability \cite{englert},
${P}$ is the predictability \cite{GY}, and $\mathcal{E}_2$ is the
concurrence, a known result due to Jakob and Bergou \cite{triality1}.

Since (\ref{triality}) is an equality for pure quanton path-detector states,
it opens up the possibility of measuring I-concurrence. In an unbalanced
$n$-path interferometer, if the quanton gets entangled with an ancilla system,
its interference visibility will be affected. Since the path predictability
is known from the experimental setting, and can also be measured, and
the interference visibility can be measured, as discussed earlier, one
can actually infer the degree of entanglelment of the quanton with the 
ancilla, without probing the ancilla in any way.

The last interesting point is that our normalized entanglement measure $\mathcal{E}$ can be expressed as the root-mean-square of \emph{concurrences} \cite{concurrence} corresponding to all pairs of paths
\begin{equation}
  \mathcal{E}^2 = \tfrac{2}{n(n-1)} \sum_{\text{pairs}}{E}_{ij}^2 ,  
\end{equation}
where ${E}_{ij}$ is the concurrence of entanglement between the quanton and the path-detector when all except path $i,j$ are blocked. This is true if all the paths are equally probable. For unequally probable paths, the above expression modifies to 
$\mathcal{E}^2 = \tfrac{n}{2(n-1)} \sum_{\text{pairs}}(\rho_{ii}+\rho_{jj})^2 {E}_{ij}^2$. This is an interesting connection between concurrence and I-concurrence which becomes meaningful in the kind of experiment we have been discussing, where pairs of paths are opened at a time.

\subsection{How quantum is the connection?}
In the light of the results obtained here, we look back at an earlier work where a triality relation has been derived for two-slit interference, in the classical optical domain \cite{triality2}. The triality relation obtained in that work is precisely of the form (\ref{triality}). It is then natural to wonder how quantum is the relation obtained here. In classical optics, there are light fields arriving on the screen from the two slits, which may differ in intensity and may have a varying degree of mutual coherence. Since the two fields differ in intensity, one can talk about a measure of particle nature which is referred to as predictability. The word distinguishability has often been used to refer to predictability in the classical optics literature, and has caused some confusion. However, there is no sense in which one can talk of distinguishability in classical optics. Distinguishability pertains to the question, which slit did the particle go through? For classical fields, it is not meaningful to ask such a question. Then if in the presence of polarization, or some other mode, the duality relation of the form (\ref{dualitygy}) becomes an inequality, it is natural to think that the duality relation is \emph{incomplete}, and can be completed by including the entanglement with polarization or some other mode. That has been the focus of the work in Ref. \cite{triality2}.

On the other hand, in the quantum regime, one can talk of path distinguishability in the context of experimentally determining through which slit did the quanton pass. This distinguishability is a quantum feature, and is intimately tied to entanglement. Thus the duality relation of the form (\ref{dualitye}), involving distinguishability is always an equality, for a pure state. So the triality relation derived here is purely quantum in this sense. It not only quantifies the connection between predictability and interference through the entanglement which may have been generated because of the quanton interacting with an external quantum system, before it arrives on the screen, it also quantitatively connects distinguishability with predictability.

Interestingly, when one moves away from single particle build up of interference, and is in the regime where the various paths may have many particles, the distinction between true entanglement and mere non-separability becomes blurred. The triality relation then continues to describe the inteference and predictability faithfully, but via classical entanglement (non-separability). In fact, for $n=2$, the triality relation obtained here is exactly the same as the one obtained in the classical optical regime \cite{triality2}, provided one identifies the absolute value of the off-diagonal density matrix element with the mutual coherence function in classical optics \cite{coherence}.

\section{Perspective}

As mentioned before, a triality relation has been obtained earlier in the context of multi-slit quantum interference \cite{triality3}. Let us look at the results obtained here in that perspective. The relation obtained earlier involved the coherence measure introduced in the context of quantum information theory \cite{plenio}. That coherence has been shown to be a proper measure of coherence, in the sense that it does not increase under any incoherent operation. On the other hand, the Hilbert-Schmidt coherence used by us, although very commonly used, has been shown to not be a robust coherence monotone, as it may increase under some incoherent operations. Nevertheless, it has been recently highlighted that in addition to having a simpler structure, the Hilbert-Schmidt coherence has information theoretic significance, in particular, through its manifestation as quantum uncertainty \cite{Luo}. Furthermore, the shortcoming of monotonicity is absent in the genuine quantum coherence framework \cite{Vicente}. Therefore, Hilbert-Schmidt coherence may still be good enough to quantify visibility in multipath interference. From a practical consideration too, the coherence measure of Baumgratz et al. has a simpler connection to multislit interference \cite{coherence}, which has also been experimentally demonstrated \cite{chen,padua}.

The predictability used in both the works has been shown to be a proper measure of path predictability. Both can be easily measured in a multipath interference experiment, as it only involves measuring the intensities at each slit, or in each path. 

The measure of entanglement is an aspect in which the earlier work falls short. The quantifier used there has not yet been shown to be a proper entanglement measure. On the other hand, the quantifier used for entanglement in the present work is I-concurrence, a well-known entanglement monotone. This advantage allows the possibility of experimentally inferring the degree of entanglement between the quanton and an ancilla system.

Thus the two formulations connecting predictability, entanglement, and interference, are equivalent, but both have their strengths and weaknesses. For two-path interference, they become identical.

\section{conclusion}

In conclusion, we have introduced an interference visibility based on the
Hilbert-Schmidt coherence, and shown that it constitutes a good measure
of wave nature, and does not suffer from the shortcomings of the conventional
fringe contrast, which emerges in certain experiments \cite{mei}. Interestingly,
this visibility can be obtained as a root mean square of the fringe contrasts
from all \emph{pairs of paths}, if the experiment is done by opening only
a pair of paths at a time. Selectively opening pairs of slit in a multi-slit
interference experiment has already been realized \cite{padua}.

A multipath predictability has been introduced, which is a natural
generalization of the two-path predictability of Greenberger and Yasin
\cite{GY}. Interestingly, this multipath predictability
can also be measured using the two-path predictabilities
of pairs of paths. A multipath distinguishability has also been introduced
which, again, can be expressed in terms of root mean square of the two-path 
distinguishability (introduced by Englert) of all pairs of paths.
The predictability and visibility introduced here, together with the
I-concurrence (related to entanglement between the quanton and a 
path-detector), are shown to follow a tight triality relation. This
brings out the essential role of entanglement in the phenomenon of
complementarity, in the general context of multipath interference. As an interesting offshoot, the normalized I-concurrence turns out to be the root mean square of concurrences relating to all path pairs.

\begin{acknowledgments}
\end{acknowledgments}
AKR, NKC and PKP gratefully acknowledge the financial support from DST, India through Grant No. DST/ICPS/QuST/Theme1/2019/2020-21/01. NP acknowledges financial support from DST, India through the Inspire Fellowship (registration number IF180414).


\begin{thebibliography}{0}%
\makeatletter
\providecommand \@ifxundefined [1]{%
 \@ifx{#1\undefined}
}%
\providecommand \@ifnum [1]{%
 \ifnum #1\expandafter \@firstoftwo
 \else \expandafter \@secondoftwo
 \fi
}%
\providecommand \@ifx [1]{%
 \ifx #1\expandafter \@firstoftwo
 \else \expandafter \@secondoftwo
 \fi
}%
\providecommand \natexlab [1]{#1}%
\providecommand \enquote  [1]{``#1''}%
\providecommand \bibnamefont  [1]{#1}%
\providecommand \bibfnamefont [1]{#1}%
\providecommand \citenamefont [1]{#1}%
\providecommand \href@noop [0]{\@secondoftwo}%
\providecommand \href [0]{\begingroup \@sanitize@url \@href}%
\providecommand \@href[1]{\@@startlink{#1}\@@href}%
\providecommand \@@href[1]{\endgroup#1\@@endlink}%
\providecommand \@sanitize@url [0]{\catcode `\\12\catcode `\$12\catcode
  `\&12\catcode `\#12\catcode `\^12\catcode `\_12\catcode `\%12\relax}%
\providecommand \@@startlink[1]{}%
\providecommand \@@endlink[0]{}%
\providecommand \url  [0]{\begingroup\@sanitize@url \@url }%
\providecommand \@url [1]{\endgroup\@href {#1}{\urlprefix }}%
\providecommand \urlprefix  [0]{URL }%
\providecommand \Eprint [0]{\href }%
\providecommand \doibase [0]{http://dx.doi.org/}%
\providecommand \selectlanguage [0]{\@gobble}%
\providecommand \bibinfo  [0]{\@secondoftwo}%
\providecommand \bibfield  [0]{\@secondoftwo}%
\providecommand \translation [1]{[#1]}%
\providecommand \BibitemOpen [0]{}%
\providecommand \bibitemStop [0]{}%
\providecommand \bibitemNoStop [0]{.\EOS\space}%
\providecommand \EOS [0]{\spacefactor3000\relax}%
\providecommand \BibitemShut  [1]{\csname bibitem#1\endcsname}%
\let\auto@bib@innerbib\@empty
\end{thebibliography}%


\begin{thebibliography}{0}

\bibitem{bohr} N. Bohr, The quantum postulate and the recent development of atomic theory.
\href{ https://doi.org/10.1038/121580a0}{\emph{Nature} (London)  \textbf{121}, 580-591 (1928).}

\bibitem{wootters} W. K. Wootters and W. H. Zurek,
Complementarity in the double-slit experiment: Quantum nonseparability and a quantitative statement of Bohr's principle.
\href{https://doi.org/10.1103/PhysRevD.19.473}{{\em Phys. Rev. D} \textbf{19}, 4
73 (1979).}

\bibitem{GY} D. M. Greenberger and A. Yasin,
Simultaneous wave and particle knowledge in a neutron interferometer.
\href{https://doi.org/10.1016/0375-9601(88)90114-4}{{\em Phys. Lett. A}  \textbf{128}, 391 (1988).}

\bibitem{jaeger} G. Jaeger, A. Shimony, and L. Vaidman, Two interferometric complementarities.
\href{https://doi.org/10.1103/PhysRevA.51.54}{\emph{Phys. Rev. A} \textbf{51}, 54 (1995).}

\bibitem{englert} B-G. Englert, Fringe visibility and which-way information: an inequality. \href{https://doi.org/10.1103/PhysRevLett.77.2154} {{\em Phys. Rev. Lett.} \textbf{77}, 2154 (1996).}

\bibitem{cd15}  M. N. Bera, T. Qureshi, M. A. Siddiqui, and A. K. Pati,
Duality of quantum coherence and path distinguishability.
\href{http://dx.doi.org/10.1103/PhysRevA.92.012118}
{\emph{Phys. Rev. A} \textbf{92}, 012118 (2015).}

\bibitem{nduality} T. Qureshi and M. A. Siddiqui, Wave-particle duality in N-path interference.
\href{http://dx.doi.org/10.1016/j.aop.2017.08.015}
{\emph{Ann. Phys.} \textbf{385}, 598-604 (2017).}

\bibitem{bagan} E. Bagan, J. Calsamiglia, J. A. Bergou, M. Hillery, Duality games and operational duality relations.
\href{https://doi.org/10.1103/PhysRevLett.120.050402}{\emph{Phys. Rev. Lett.} \textbf{120}, 050402 (2018).}

\bibitem{plenio} T. Baumgratz, M. Cramer, M. B. Plenio, Quantifying Coherence.
\href{https://doi.org/10.1103/PhysRevLett.113.140401}
{\emph{Phys. Rev. Lett.} \textbf{113}, 140401 (2014).}

\bibitem{coherence} T. Qureshi, Coherence, interference and visibility.
\href{http://dx.doi.org/10.12743/quanta.v8i1.87}
{\emph{Quanta} \textbf{8}, 24 (2019).}

\bibitem{roy} P. Roy and T. Qureshi,  
Path predictability and quantum coherence in multi-slit interference.
\href{http://dx.doi.org/10.1088/1402-4896/ab1cd4} 
{\emph{Phys. Scr.} \textbf{94}, 095004 (2019).}

\bibitem{triality1} M. Jakob and J. A. Bergou, Quantitative complementarity relations in bipartite systems: Entanglement
as a physical reality.
\href{https://doi.org/10.1016/j.optcom.2009.10.044}
{\emph{Optics Communications} \textbf{283}, 827 (2010).}

\bibitem{concurrence} S. Hill and W. K. Wootters, Entanglement of a pair of quantum bits.
\href{https://doi.org/10.1103/PhysRevLett.78.5022}
{\emph{Phys. Rev. Lett.} \textbf{78}, 5022 (1997).}

\bibitem{triality2} X.-F. Qian, A. N. Vamivakas, and J. H. Eberly, 
Entanglement limits duality and vice versa.
\href{https://doi.org/10.1364/OPTICA.5.000942}
{\emph{Optica} \textbf{5}, 942 (2018).}

\bibitem{triality3} T. Qureshi, Predictability, distinguishability, and entanglement.
\href{https://doi.org/10.1364/OL.415556}
{\emph{Opt. Lett.} \textbf{46}, 492 (2021).}

\bibitem{huber} B. C. Hiesmayr, M. Huber, Multipartite entanglement measure for all discrete systems.
\href{https://doi.org/10.1103/PhysRevA.78.012342}
{\emph{Phys. Rev. A} \textbf{78}, 012342 (2008).}

\bibitem{rungta} P. Rungta, V. Buzek, C. M. Caves, M. Hillery, G. J. Milburn, Universal state inversion and concurrence in arbitrary dimensions.
\href{http://dx.doi.org/10.1103/PhysRevA.64.042315}
{\emph{Phys. Rev. A} \textbf{64}, 042315 (2001).}



\bibitem{mei} M. Mei and M. Weitz, Controlled decoherence in multiple beam Ramsey interference. 
\href{https://doi.org/10.1103/PhysRevLett.86.559}
{\emph{Phys. Rev. Lett.} \textbf{86}, 559 (2001).}

\bibitem{luis} A. Luis, Complementarity in multiple beam interference.
\href{https://doi.org/10.1088/0305-4470/34/41/314}
{\emph{J. Phys. A: Math. Gen.} \textbf{34}, 8597 (2001).}

\bibitem{bimonte1} G. Bimonte and R. Musto, On interferometric duality in multibeam experiments.
\href{https://doi.org/10.1088/0305-4470/36/45/009}
{\emph{J. Phys. A: Math. Gen.} \textbf{36}, 11481–11502 (2003).}

\bibitem{prillwitz} K. von Prillwitz, L. Rudnicki, F. Mintert, Contrast in multi-path interference and quantum coherence.
\href{https://doi.org/10.1103/PhysRevA.92.052114}
{\emph{Phys. Rev. A} \textbf{92}, 052114 (2015).}


\bibitem{durr} S. D\"{u}rr, Quantitative wave-particle duality in multibeam interferometers.
\href{https://doi.org/10.1103/PhysRevA.64.042113}
{\emph{Phys. Rev. A} \textbf{64}, 042113 (2001).}

\bibitem{Luo} Y. Sun and S. Luo, Coherence as uncertainty.
\href{https://doi.org/10.1103/PhysRevA.103.042423}
{\emph{Phys. Rev. A} \textbf{103}, 042423 (2021).}

\bibitem{Vicente} J. I de Vicente and A. Streltsov, Genuine quantum coherence.
\href{https://doi.org/10.1088/1751-8121/50/4/045301}
{\emph{J. Phys. A: Math. Theor.} \textbf{50}, 045301
 (2017).}
 
\bibitem{chen} X. Chen, Y. Deng, S. Liu et al.,
A generalized multipath delayed-choice experiment on a large-scale quantum nanophotonic chip.
\href{https://doi.org/10.1038/s41467-021-22887-6}
{\emph{Nat. Commun.} \textbf{12}, 2712 (2021).}

\bibitem{padua} P. Machado and S. Pádua,
Quantum coherence of spatial photonic qudits: experimental
measurement and path-marker analysis. 
\href{https://doi.org/10.1088/2040-8986/ab8451}
{\emph{J. Opt.} \textbf{22} 065201 (2020).}

\bibitem{helstrom} C. W. Helstrom, Quantum detection and estimation theory (Academic Press, New York, 1976).

\bibitem{gconc} V. S. Bhaskara and P. K. Panigrahi, 
\href{http://dx.doi.org/10.1007/s11128-017-1568-0}
{\emph{Quantum Inf. Process.} \textbf{16}, 118 (2017)}; A. K. Roy, N. K. Chandra, S. N. Swain, and P. K. Panigrahi, \href{https://doi.org/10.1140/epjp/s13360-021-02127-y}
{\emph{Eur. Phys. J. Plus} \textbf{136}, 1113 (2021).}

\bibitem{bimonte2} G. Bimonte and R. Musto, Comment on `Quantitative wave-particle duality in multibeam interferometers'.
\href{https://doi.org/10.1103/PhysRevA.67.066101}
{\emph{Phys. Rev. A} \textbf{67}, 066101 (2003).}

\end{thebibliography}
\end{document}